# Dynamic Interlining in Bus Operations

Seyedmostafa Zahedi[1], Haris N. Koutsopoulos[1], Zhenliang Ma[2]
[1] *Department of Civil and Environmental Engineering, Northeastern University, Boston, MA 02115, United States*
[2] *Department of Civil and Architectural Engineering, KTH Royal Institute of Technology, Stockholm 11428, Sweden*

**ABSTRACT**

The paper introduces and evaluates the concept of the dynamic interlining of buses. Dynamic interlining is an operational strategy for routes that have a terminal station at a common hub, that allows a portion of (or all) the fleet to be shared among the routes belonging to the hub (shared fleet) as needed. The shared fleet is dispatched on an on-demand basis to serve scheduled trips on any route to avoid delays and regulate services. The paper examines systematically the impacts of dynamic interlining on service reliability. It formulates the dispatching problem as an optimization problem and uses simulation to evaluate the dynamic interlining strategy under a variety of operating conditions. Using bus routes in Boston's Massachusetts Bay Transportation Authority (MBTA) as a case study, the feasibility of the strategy, as well as factors that affect its performance are investigated. Results show that dynamic interlining can improve service reliability (increases on-time departures and decreases departure headways variability at the hub). The fraction of the fleet that is shared has the most dominant impact on performance. In the case where all buses are dynamically interlined, the performance improves as route frequency increases and more routes participate in the strategy.

## 1. Introduction

Bus routes are often designed along corridors that consolidate individual passengers' needs for service. The aggregation provides economies of scale, serving as many passengers as possible, and increasing efficiency. However, it comes at the cost of having a rigid service that is not inherently responsive to emerging conditions in demand patterns and operating environments. One key aspect of service quality is service reliability, which is the delivery of the service as scheduled (Lai and Chen, 2011). Service reliability is challenging as transit operations are stochastic in nature and run times are random (Sánchez-Martínez, 2012). To overcome the uncertainties associated with operations, services are often scheduled assuming near worst-case scenarios. For example, for determining the fleet size of a bus route, planners use values of higher percentiles (commonly $85^{th}$ - $95^{th}$) of run time distributions to decrease departure delays due to the late arrival of a bus (Sánchez-Martínez et al., 2016). However, the gains in reliability take place at the expense of the increased fleet size required to operate the resulting schedule and, consequently, reduced operating efficiency. The lower efficiency often manifests itself in buses remaining idle at end stations or drivers intentionally slowing down or holding at stops to avoid being early (Sindzingre, 2019).

Adding flexibility to bus services and operations has the potential to improve service quality and reliability by adjusting to travel demand, as well as increase productivity and efficiency (fleet utilization) during times of scarce resources and limited budgets. Flexible transit services have been postulated in the literature and experimented with, by transit agencies, as means to improve the level of service and attractiveness (Alshalalfah, 2009). A guide to plan and operate such services as well as best practices for their success are discussed in (Potts et al., 2010).

Transit can be flexible with respect to the service provided to users and with respect to the operations (how the service is delivered). Services can be flexible in their routes, schedules communicated to the riders, or the mode options provided to the customers. The flexibility in services is visible to the rider. The flexibility in operations is from the perspective of the operator and it pertains to fleet management and strategies deployed to deliver a given service (Figure 1).

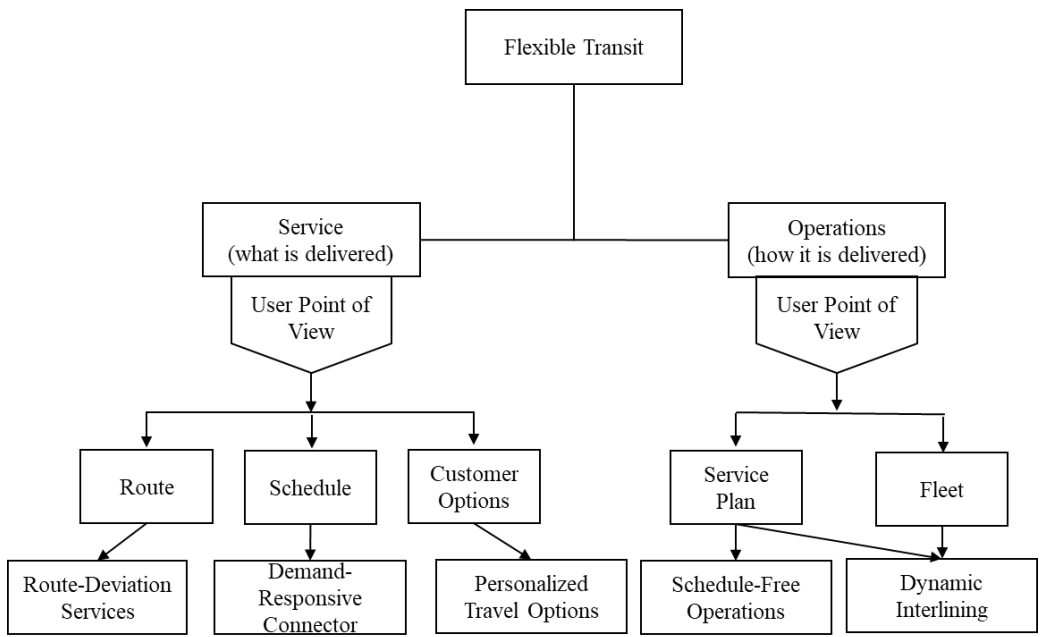

**Figure 1. Design aspects of flexible transit**

On the service side, Koffman (2004)discusses various types of transit services that allow some degree of deviations from a planned route. The deviations are limited to a buffer area around the actual route and are made possible by introducing slack times in the bus schedules. The bus follows an overall path and has to meet a schedule at specific checkpoints along the route. However, it can deviate and serve curb-to curb requests between the checkpoints. Point deviation services are another example that varies in terms of degrees of flexibility en-route and the need to operate on a defined path (Zheng et al., 2018). Flexibility in schedules is used in the operations of some Demand Responsive Connectors (DRC) (De almedia and Felipe, 2019). DRCs provide first mile/last mile services to passengers connecting to a major transit terminal on an on-demand basis. There are many variations of DRCs with first and last mile the most common (Yang et al., 2021). In the first mile mode, passengers who want to use the transit terminal are collected from their origins and taken to the terminal. In the last mile mode, passengers who have arrived at the terminal are picked up and taken to their final destinations. Depending on the number of passengers requesting service, the DRC schedule can change. Flexibility can also be viewed as providing different modes of transit. Passengers can choose different modes based on emerging conditions. Atasoy et al. (2015) introduced a flexible system in which a list of transit options is provided to passengers who have the flexibility to choose from a menu of personalized services.

From the perspective of operating flexibility, which is of most interest to this paper, Sánchez-Martínez et al. ( 2017) proposed a schedule free paradigm for high frequency transit operations in which buses do not have specific arrival/departure times. Schedule-free transit is driven by optimizing in real-time the operating plans that govern future trips of each vehicle by specifying the sequence of stops and their target arrival and departure. The plan is updated in real time based on emerging conditions, and can include strategies such as holding, stop-skipping, short-turning, deadheading, expressing, and injecting reserve vehicles. Similar to this study, Furth and Nash (1985) proposed a vehicle pooling strategy, which does not dedicate buses to a specific route but uses the buses on a first come, first served basis among all the routes. In this strategy, buses that serve routes emanating from a common terminal are not dedicated to any route and are not assigned to specific trips either. Therefore, they can serve any trips that leave the terminal.

Other than the design aspects, many real time operational strategies have been proposed to improve reliability. Vehicle holding is a common method used to regulate the service and prevent bunching (Laskaris et al., 2019). Different holding strategies can be deployed based on the route frequency. Low frequency routes use schedule-based holding which aims at having buses depart from checkpoints according to schedule. At higher frequency routes, headway-based holding is common to regulate bus departures from checkpoints relative to the previous departure. Other real-time control strategies to improve service reliability, such as stop skipping, short turning, and signal preemption, have also been proposed by Zhang et al. (2017) and Turnquist (1974).

Building on Sánchez-Martínez et al. (2017) and Furth and Nash (1985) the paper introduces the concept of dynamic interlining as means to increase flexibility in the operations of bus services and provide reliable service with reduced resources compared to current practice. Traditionally, interlining is the practice of scheduling buses to operate trips on different routes (referred to as static interlining in the paper). Static interlining is a popular practice among agencies with the objective of reducing fleet size. Lack of layover locations, and eliminating end-of-line looping are other reasons for static interlining (Pine et al., 1998). Dynamic interlining resembles static interlining in that buses can be used on different routes. However, it is different from the static interlining in several aspects, for example, the switch from one route to another is planned in advance in static interlining, while the switch is decided in real time in dynamic interlining.

In Dynamic interlining, instead of dedicating a predetermined number of buses to each route, calculated using a high percentile of run times, buses (or at least a portion of them) are shared among the routes in a dynamic manner, depending on emerging conditions. When a shared bus enters the hub, instead of being dispatched only on a specific route at a specific time, it can be dispatched on any of the routes in the hub. The structure of many transit networks is characterized by a number of hubs with routes emanating from them, therefore, there is potential to reduce fleet size without sacrificing (or even exceeding) reliability by dynamically interlining the buses.

The paper aims to design functions required to support dynamic interlining, systematically evaluate the potential of dynamic interlining to improve service reliability, and identify key factors that affect its performance. The concept is similar to the one discussed in Furth and Nash (1985), However, the analysis is limited in several ways. There was no analysis of a mixed fleet of dedicated and shared buses or the impacts of different dispatching strategies and the characteristics of bus routes that can benefit from the strategy. This paper generalizes the concept proposed in Furth and Nash (1985). It also proposes a method for optimally deciding when and where buses should be dispatched. The main contributions of the paper are as follows:
a) Formulation of the dynamic interlining problem and theoretical justification of its potential.
b) Formulation and solution of the optimal bus dispatching in dynamic interlining problem.
c) Systematic evaluation of the concept addressing the following questions:
- Can dynamic interlining help reduce the fleet size?
- How should the shared fleet be used by the participating routes?
- What are the characteristics of routes and the configuration of the (e.g. number of routes) that are favorable to adopting dynamic interlining?

The rest of the paper is organized as follows. Section 2 defines the dynamic interlining strategy and introduces the notation used in the paper. Section 3 uses analytical approximations to examine the potential of the strategy. Section 4 formulates the problem of how shared vehicles are assigned to routes as an optimization problem. Section 5 presents the methodology to evaluate dynamic interlining in detail. Section 6 discusses the application of dynamic interlining using a set of bus routes from the Massachusetts Bay Transportation Authority's (MBTA) network and investigates the merits of the idea through a number of experiments. Section 7 concludes the paper.

## 2. Problem Description

Traditional transit scheduling assigns a fixed fleet of buses to a route that either serves only that route or is statically interlined with other routes. Transit agencies sometimes prefer to minimize the use of static interlinings as they tend to propagate delays from one route to another. Alternatively, buses are dedicated to a route and dispatched at a scheduled departure time or at a certain frequency.

If all vehicles are used exclusively on a route (no static interlining), the number of vehicles $N$ required to operate a service with a target level of reliability depends on the design cycle length $C$ and headway $h$, and can be estimated using Equation 1 (Pine et al., 1998).

$$N = \left\lceil C * \frac{1}{h} \right\rceil \tag{1}$$

Where, $N$ is the required fleet size, $C$ is the cycle time, and $h$ is the headway, and $\lceil . \rceil$ is the round up operator. Cycle time is the time needed to make a round trip on the route, including layover/recovery time and is usually determined as a (high) percentile of the distribution of bus run times (Sánchez-Martínez, 2012).

As already mentioned, many transit networks consist of a number of hubs with routes starting and ending at the hubs. Let us consider a part of a transit network as shown in Figure 2. There are two hubs and a number of routes. Often, each route has its own fleet and operates independently from other routes. Considering that cycle times are defined as a high percentile of run times, buses may require less time than scheduled to arrive at the terminal. Dynamic interlining allows some of the buses that may be idle to be dispatched, if conditions allow, on other routes, instead of waiting for a specific departure time on a dedicated one, and used as shared buses. The motivation is that the strategy can lead to similar performance but with a reduced fleet size, since buses can be utilized more efficiently when shared. The example in Figure 2 illustrates the concept. Four buses are present at the hub locations where they can be shared by all routes. If a bus is late and cannot execute its next scheduled departure, then a shared bus in the "pool", if available, would be dispatched instead. In this manner, the shared buses can be dispatched as needed, while increasing the chances of meeting the schedule for the different routes.

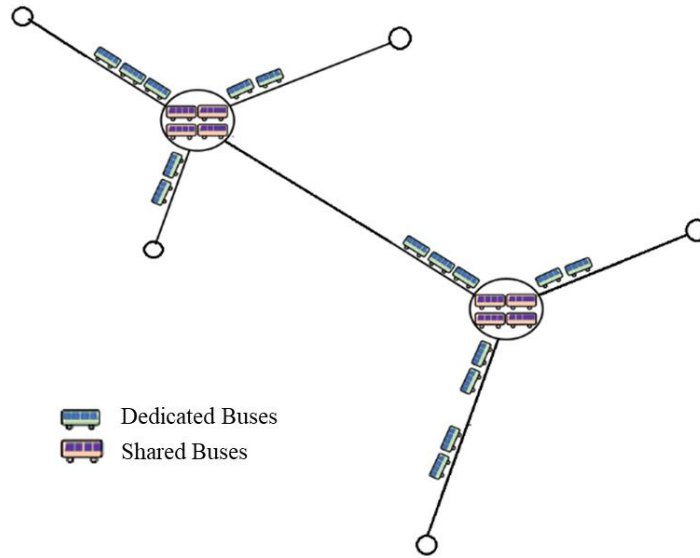

**Figure 2. Dynamic interlining and shared fleet**

If a portion of the buses is dynamically interlined, none of the buses have schedules i.e. they are not assigned to specific trips. Although buses could still be dedicated to a specific route, within each route buses are flexible to cover trips for each other. Once a bus from the shared fleet is dispatched instead of a dedicated bus, the dedicated bus that is late should be available to do any trip for the route to which it belongs. A special case of dynamic interlining with no shared buses is explored in section 3.1.

We now define more formally the dynamic interlining problem. Let us consider a hub-and-spoke bus network, with a set $R$ of bus routes $r \in R$, starting from/ending at the hub. Each route has a pre-specified scheduled headway $h_r^s$, and stochastic round trip travel time $t_r$. The bus fleet is mixed and is represented by $B$. The number of buses dedicated to route $r$ is $B_r$ and the number of buses that can be shared on as needed basis is $B_S$. The problem is defined as:

*Given the hub-and-spoke bus network and service characteristics, optimize the fleet size allocation and dispatching strategy of the dedicated and shared buses in order to satisfy the service requirements (schedule/frequency) while providing a target level of service reliability (e.g. on-time performance).*

Some important practical considerations of dynamic interlining include:
A. *Idle buses and hub capacity.* Limited bus storage capacity at hubs can hinder operators from adopting dynamic interlining. The notion of reserving a portion of the fleet to be used on an on-demand basis may suggest that extra space at the hubs may be needed. The average number of buses that are idle at a hub is a function of the average run time distribution, fleet size, and average frequency. We expect that under dynamic interlining, buses are better utilized which lead to lower idle times. Furthermore, we expect that for the same level of service a smaller fleet is needed, which will reduce idle times as well. For further evidence, Section 3 discusses the problem using a queuing theory analogy. In Section 6 the results show that dynamic interlining actually reduces average bus idle times.
B. *Within-hub travel times.* Often the terminals of the various routes, although associated with the same hub, may not be located at the same point. We define, the within-hub travel time as the time it takes for a bus to travel from the end point of the route when entering the hub to the start point of the next route that is assigned to for the next trip. These "set up" times should be considered in the analysis, and they may impact the effectiveness of the dynamic interlining strategy. The empirical analysis in section 6 investigates the impact of the set up times on performance.
C. *Real time control of shared buses.* Dynamic interlining controls the dispatching of shared buses from terminals. Once the buses are dispatched from the terminal, other strategies can be employed to regulate the service along the route. Bus bunching, for example, happens when a bus catches up with the bus in front (Pilachowski, 2009), (Feng and Figliozzi, 2011). Some operators allow the buses to pass each other, while others do not. The decision to allow buses to pass each other or not may have consequences on the performance of dynamic (investigated in Section 6).
D. *Crew scheduling and work rules.* Dynamic interlining may require drivers to operate on all participating routes. Also, since the trips each driver takes are not scheduled in advance, it may happen that a driver's last trip is on a long route resulting in an overtime payment to that driver. In addition, some union contracts require drivers to finish their last trips where they started. Such constraints complicate the operation of dynamic interlining and are out the scope of the paper since the paper's main objective is to demonstrate the potential of dynamic interlining. In any case, it is possible to add constraints that require the last trip of a driver to be on a specific route.

Table 1 summarizes the notation that is used throughout the paper.

**Table 1. Notation**

| | |
|---|---|
| $R$ | *Set of all routes from the hub* |
| $B$ | *Set of all buses under consideration* |
| $B_r^h$ | *Set of dedicated buses available at the hub for route r* |
| $B_s^h$ | *Set of shared buses available at/arriving to the hub* |
| $B_r$ | *Set of k nearest buses on route r on their way to the hub* |
| $D$ | *Set of all trips under consideration* |
| $D_r$ | *Set of immediate trips of route r* |
| $\mathcal{M}$ | *Set of all potential assignments of buses to trips* |
| $w_q$ | *Wait time in queue* |
| $C_a$ | *Coefficient of variation of inter-arrival times* |
| $C_s$ | *Coefficient of variation of service times* |
| $\lambda$ | Arrival rate |
| $\mu$ | Service rate |
| $D_{r,i}$ | The $i^{th}$ scheduled departure on route $r$ |
| $T$ | Time period under consideration |
| $t_r$ | Run time of a trip on route $r$ (random variable) |
| $t_r^i$ | Run time of a bus serving the $i^{th}$ trip on route $r$ |
| $t_{r,i}$ | Run time of bus $i$ on route $r$ |
| $t_{t,r}^{sd}$ | Scheduled departure time for trip $t$ on route $r$ |
| $d_i$ | Departure delay for trip $i$ (random variable) |
| $\beta$ | Intercept for the trip level dwell time model |
| $t_{r,i}^{dewll}$ | Trip level dwell time of trip $i$ on route $r$ |
| $C$ | Cycle length |
| $f(t_r)$ | Run time distribution of route $r$ |
| $N_r$ | Size of the dedicated fleet for route $r$ |
| $S_r$ | Service plan for route $r$ |
| $F_s$ | Shared fleet size |
| $p_r^e$ | Effective percentile of route $r$ |
| $h_r^s$ | Scheduled headway of route $r$ |
| $h_i^r$ | Departure headway between trips $i$ and $i-1$ on route $r$ |
| $Q_r$ | Ridership of route $r$ (pax/hr) |
| $n_r$ | Number of trips made on route $r$ during time period $T$ |
| $N_h^b$ | The expected number of busy buses, for all routes associated with the hub |
| $P_j$ | Penalty of not assigning buses to trip $j$ |
| $C_{ij}$ | Cost of assigning bus $i$ to trip $j$ |
| $m$ | Number of routes in a hub |
| $\alpha_r^{dwell}$ | The average boarding time per passenger on route $r$ |
| $d\_f(S,t)$ | Deficit value for a terminal with schedule S as a function of t |

## 3. Potential of Dynamic Interlining. Theoretical Evidence

Dynamic interlining has the potential to improve reliability and reduce fleet size. In this section, we provide an analytical justification using a queue model approximation to assess the performance of dynamic interlining when all buses are shared among routes.

We model the operations of a route as a queuing system, in which buses are servers and the scheduled trips are the customers. The service time of a customer (trip) associated with route $r$ follows the bus run time distribution for route $r$. Customers (trips) arrive in a deterministic manner according to the schedule (e.g., every $h_r$ minutes). Waiting times in the queue are equivalent to departure delays of buses at terminals. The probability of the event that the system cannot serve a trip at its scheduled time is equal to the probability that all servers (buses) are busy.

We consider the case of a hub serving $m$ routes with identical characteristics (frequency, runtimes, etc). We assume two operating scenarios: a) each route is served by its own fleet of dedicated buses, c, and b) all buses (i.e., $m * c$ buses) are shared among the routes on a first-come-first-served basis, i.e., a trip can be served by the first available server (bus). In the first scenario, each route $r$ is modeled as a queueing system with $c$ servers, arrival rate of $1/h$ and average service time, $E(t_r) = E(t)$, the same for all routes. The arrival rate is deterministic (i.e., the schedule) and the service times follow the probability density function of the bus run times. Hence, a $D/G/c$ queueing system (Figure 3a) is used to model the operations for each route. In the second case, there is one queuing system consisting of $m*c$ servers (the entire fleet used by all routes as Figure 3b shows). The customers represent all trips scheduled for all routes. Hence, the arrival rate is the sum of arrival rates for all routes. The system is now modeled as a $D/G/m*c$ queuing system with $m*c$ servers and $m/h$ arrival rate. In both scenarios, a trip is not assigned to a specific bus (server) in advance. Rather, a trip is served by the first available bus (server).

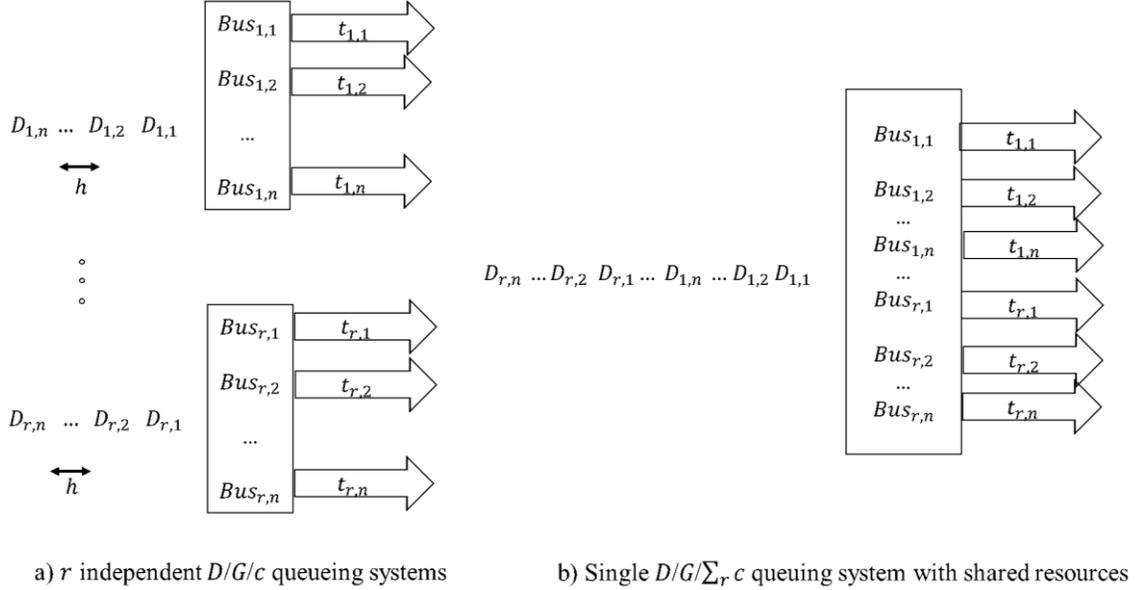

**Figure 3. Queue representation of bus routes**

Based on (Chaves, 2016), the performance of a $G/G/c$ queuing system can be approximated by Equation (2). Where the average wait time in a $G/G/c$ queue system is approximated by an $M/M/c$ queuing system and corrected for the coefficients of variation in service and inter-arrival times.

$$W_q^{G/G/c} \approx W_q^{M/M/c} \frac{C_a^2+C_s^2}{2} \qquad (2)$$

where, $W_q^{G/G/c}$ is the average wait time of a $G/G/c$ system, $W_q^{M/M/c}$ the average wait time of a $M/M/c$ system. $C_a$ and $C_s$ are the coefficients of variations of inter-arrival and service times, respectively.

Using Equation (2) we can quantify the reliability benefits of dynamic interlining and find the factors that impact its performance. Let us assume four identical routes with 12 buses per route an average run time of 1 hour, and headways of 6 minutes. The route run times follow a general distribution with a coefficient of variation (CoV) of 0.15. Consequently, in the equivalent queue representation of the problem, the service rate per server (bus), $\mu$, is 1 customer per hour and the arrival rate, $\lambda$, is 10 customers (trips) per hour.

Equation (2) can be used to estimate the average wait time in the queue (equivalent to the average departure delay) for the two operating scenarios, routes operating independently with dedicated buses or with dynamically interlining shared resources. $C_a$, the CoV of inter-arrival times (scheduled) is 0 and $C_b$, the CoV of service times (run times) is 0.15. The results show that a single route with 12 buses (utilization ratio of 0.83) has an average departure delay of 9.1 seconds. If all four routes share their buses, the utilization ratio remains the same, but the average departure delay decreases to 0.8 seconds. If the total fleet size (number of servers) in the shared case is reduced from 48 to 43 the average departure delay is 7.8 sec. This is lower than the 9.1 seconds expected in the case with dedicated buses and the full fleet (12 buses per route). The results indicate that the dynamic interlining of shared buses has the potential to save resources and improve reliability at the same time.

The queue representation of the system allows the investigation of other factors that may impact the performance of dynamic interlining, such as run time variability, and the number of routes that are dynamically interlined. In practice, larger variations in run times, require larger cycle times to maintain service reliability. Therefore, it is expected that the benefits of dynamic interlining will increase with run time variability. Figure 4 shows the average departure delays for operating 4 bus routes with dedicated buses compared to the case with shared resources under different run time variabilities. The routes have the same characteristics in terms of average run times, headways, and fleet size. In order to explore the effects of run time variability, the cycle length and fleet size of the routes are kept the same. The results clearly show that, as the variation in run times increases, the benefits of dynamic interlining (sharing the buses) increase.

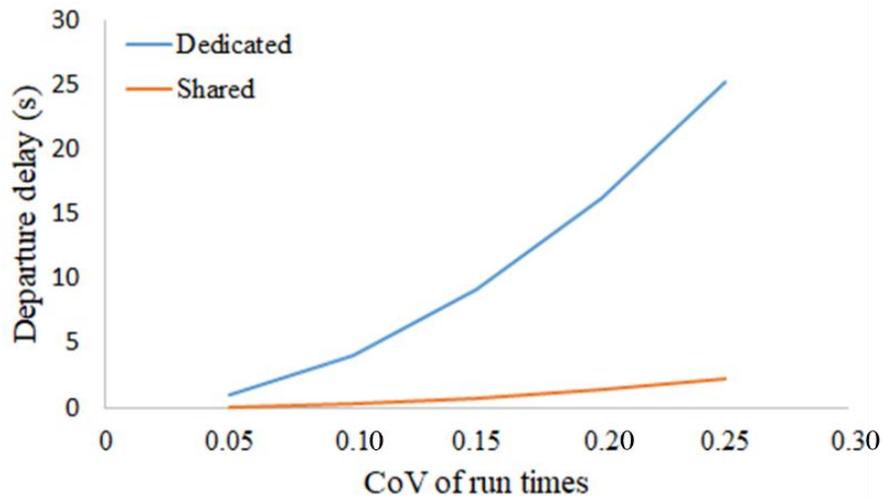

**Figure 4. Impact of run time variability on the performance of dynamic interlining**

The dynamic interlining works by sharing resources among routes. Therefore, its performance should improve as more routes share vehicles. Figure 5 shows the average departure delay for a different number of routes participating in dynamic interlining as a function of the utilization ratio. The utilization ratio in this case is the ratio of the number of trips per hour ($\lambda$) to the number of trips per hour that the fleet can perform ($\mu$). The results indicate that for a given utilization ratio the departure delay decreases as more routes share resources. That means for a given performance target (i.e. delay), the required fleet size per route decreases as the number of routes involved in dynamic interlining increases (higher utilization ratios can be used).

The queuing theory analogy also supports that the idle time is lower in the dynamic interlining case, as buses are better utilized. The idle time is equivalent to the probability that the server system is idle in the system and this probability is higher in the case of $n\ M/M/c$ queuing systems compared to an $M/M/nc$ queuing system (analytical results for the probability of the system being idle exists for $M/M/c$ systems but not for $G/G/c$ systems; however, the underlying principle is the same).

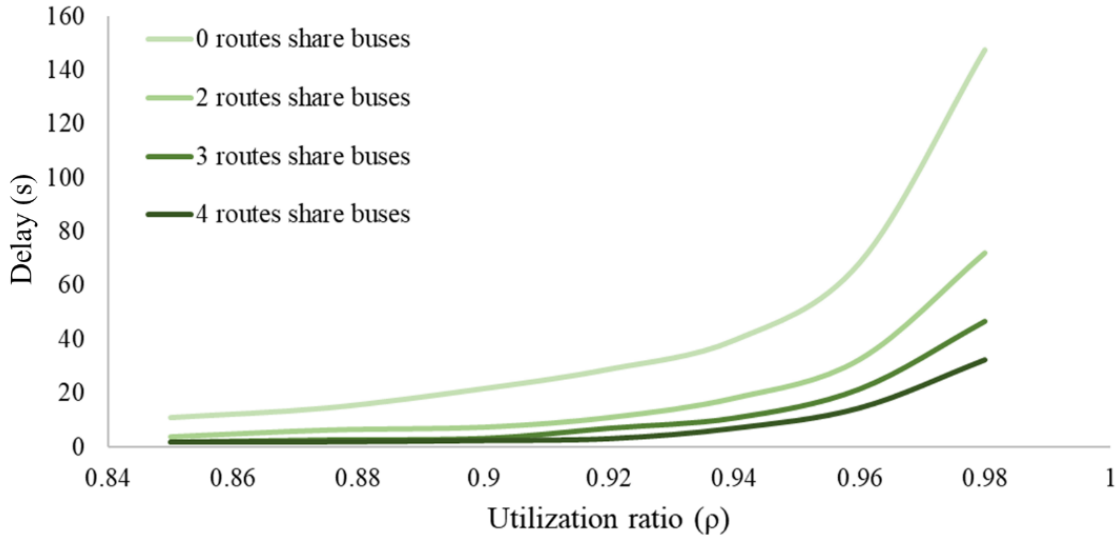

**Figure 5. Impact of the number of routes on the performance of dynamic interlining (0 routes share buses means no interlining and each route operates independently).**

## 4. DECISION SUPPORT FOR DYNAMIC INTERLINING

The queuing approximation of the problem provides insights into the potential of the dynamic interlining strategy to improve service reliability and how various route characteristics impact its performance. However, it is a high-level abstraction of bus operations and the analysis is limited by assumptions on route attributes. For example, all bus routes are assumed to have the same run time distribution, otherwise, the bus service time will depend on the route which violates the required assumption that service times are independent and identically distributed In addition, the queue model assumed an all-or-nothing approach. That is, it only allows comparison of scenarios in which either all buses are shared, or all are dedicated to their respective routes. However, strategies, where only a fraction of the buses are shared, are also of interest, for example, due to constraints such as driver familiarity with routes, vehicle and infrastructure configurations, etc.

A decision support system for evaluating and deploying the dynamic interlining strategy requires two major components:

1. *Fleet allocation*. A planning component that deals with decisions with respect to fleet sizes of each route as well as the size of the shared fleet. It addresses the problem of allocating resources to the shared and dedicated fleets as well as the distribution of the dedicated fleet among the routes. It aims at finding an equitable distribution of the dedicated fleet so that the performance of one route is not improved at the expense of another
2. *Dispatching strategies*. A real-time component that governs the use of the shared fleet with the aim of regulating the service and improving service reliability. It informs decisions on the use of shared resources during the operations. The decision entails dispatching a shared bus on a route that is experiencing delays or reserving the shared bus for future trips. The optimal dispatching algorithm considers the availability of the shared resources as well as the status of each route to decide if a route can use a shared bus.

## 4.1 Fleet allocation

The required fleet size to serve a route depends on the service schedule, bus run times, and targeted service reliability. Assuming constant headways, the fleet size can be calculated using Equation 1. For the general case of scheduled departures with uneven headways, the deficit function approach was proposed to estimate the required fleet size (Ceder, 2005). The deficit function is a step function measured at a particular bus terminal. It calculates the minimum number of buses required to implement a schedule $S$ during time period $T$ given run times between terminals. The required fleet size is the maximum value of the deficit function over all times in T.

$$N = \max_{t \in T} d\_f(S, t) \tag{3}$$

where $N$ is the required fleet size. $d\_f(S, t)$ is the deficit at time $t$ measured as the difference between the number of buses scheduled to depart from the hub and the number of buses scheduled to arrive at the hub up to time $t$.

The fleet size allocation problem is further complicated by dynamic interlining since a portion of the fleet (the shared fleet) is used by all routes, and the rest is dedicated to specific routes. The shared fleet size is a design parameter to be investigated. Given a shared fleet size, it is important to maintain an equitable distribution of resources among the routes in terms of dedicated buses. Figure 6 shows the approach used to allocate dedicated buses among routes.

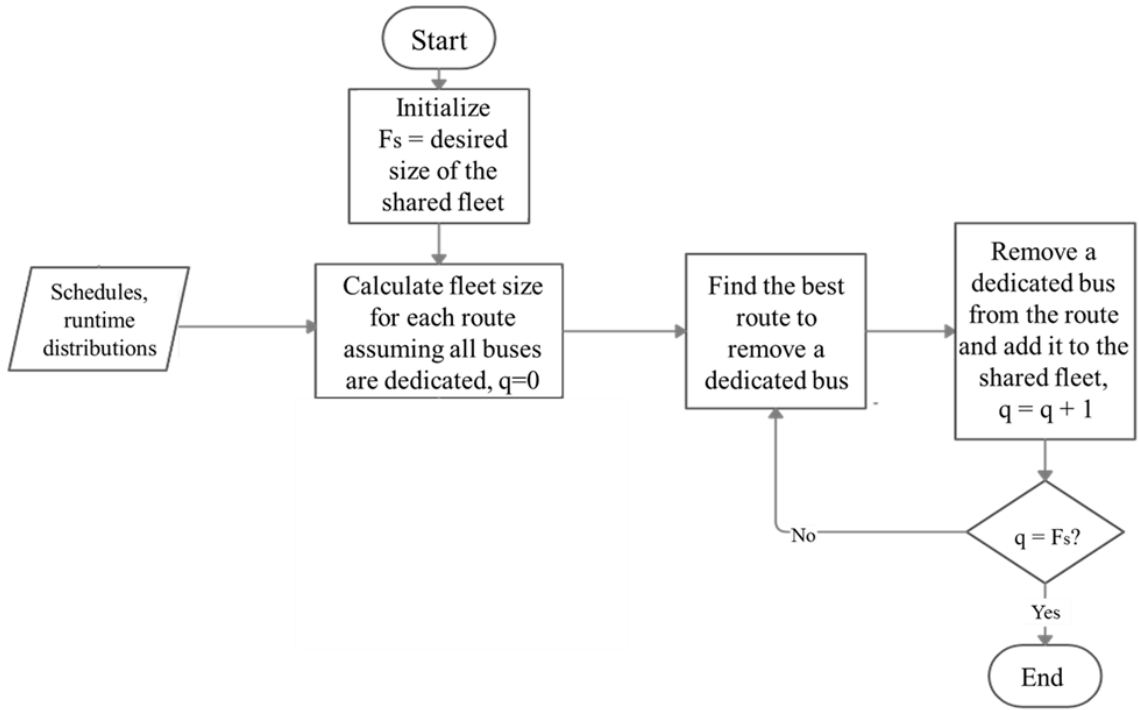

**Figure 6. Flowchart of the fleet allocation algorithm**

It consists of the following steps:

- STEP 1: Determine a base case fleet allocation for the routes assuming all buses are dedicated.

- STEP 2: Remove one bus at a time from the dedicated fleet and add it to the shared fleet. The bus that is added to the shared fleet is taken from the route whose performance is least impacted by removing that bus.
- STEP 3: Repeat step 2 until the desired size ($F_s$) for the shared fleet is reached.

Determining the route whose performance will be affected the least is not trivial. (Sánchez-Martínez et al., 2016) used a simulation-based optimization approach to allocate resources among a group of independent routes aiming at maximizing their collective performance. They used a greedy heuristic that compares the marginal reliability benefits of adding a bus to each route with the dis-benefit of taking one away. The bus is taken from the route whose reliability will be less compromised (marginal dis-benefit is the smallest) and allocated to the route that would benefit the most (marginal benefit is the largest).

The approach for the fleet allocation problem proposed in this paper follows the idea in Sánchez-Martínez et al. (2016). Equation 1 or 3 is used for the initial solution. In Step 2, the route that will contribute a bus to the shared fleet is determined as the one with the highest "effective percentile" of the run time distribution among all routes in the absence of one bus. The effective percentile is defined as the percentile of a route's run time distribution that corresponds to the cycle length used in Equation 1 or 3 assuming the fleet size for the route, $N_r$, has been reduced by one. For a route with a fleet size $N_r$, determined using Equation 1, the effective percentile is calculated using the following equations.

$$C_r = N_r * h_r \tag{4}$$

$$p_r^e = \int_0^{C_r} f(t_r)\, df \tag{5}$$

where $C_r$ is the new cycle length that has to be used to provide service with headway $h_r$ and reduced fleet size $N_r$, $f(t_r)$ is the probability density function of the run times of route $r$, and $p_r^e$ is the effective percentile of route $r$ with fleet size $N_r$.

At each step, the route with the largest effective percentile contributes one bus to the shared fleet and has its dedicated fleet size reduced by one. The pseudo code for the fleet allocation is:

*Bus Allocation Algorithm*

1. Initialization
   $F_s$ = desired shared fleet size, $q = 0$ (number of buses assigned to the shared fleet)
2. For every r ∈ R:
   Solve Equation 1 or 3 with service plan $S_r$ to obtain $N_r$
3. For every r ∈ R:
   Calculate the effective percentile of route r, $p_r^e$, with $N_r-1$, $f(t_r)$, and $S_r$ using Equation 5
4. r*: $\max_r \{p_r^e\}$
   $N_{r*} = N_{r*}-1$
   q = q +1
5. If q < $F_s$ go to Step 3
6. Output $N_r\ \forall\ r \in R$

## 4.2 Dispatching strategy

The dispatching policies that govern the use of the shared and the dedicated fleet can impact the performance of dynamic interlining. In operations, a dispatched shared bus will be unavailable to other routes until it returns back to the hub. The unavailability of the shared bus may impact the performance of other routes at the hub since they could have used the bus to avoid departure delays. The dispatching strategies focus on decisions regarding when and which route to use a shared bus.

The simplest strategy is to use the shared fleet on a first-come-first-serve (FCFS) basis, meaning that the shared buses are dispatched on the route that needs the shared bus sooner. The FCFS strategy can be problematic, especially when the shared fleet is small. To illustrate this, we consider the case of two routes (routes 1 and 2) with a common hub an only one shared bus. Let us assume that at the time of a trip on route 1, none of route 1's dedicated buses are available. Based on the FCFS strategy, the dispatcher will assign the shared bus to route 1. Shortly after the shared bus is dispatched, a dedicated bus for route 1 returns to the hub and now has to stay idle because its trip is covered by the shared bus. If route 2 also experiences delays, there is no shared bus available to be dispatched on route 2, even if the delay is larger than what route 1 experienced.

Another strategy is to prioritize the routes that have the largest delays. At a decision point, routes are ranked based on the delays that they are experiencing and the available shared buses are assigned to the routes in a descending order of delay until either the shared fleet is depleted or no route needs to use a shared bus. A minimum delay threshold can be used to save the shared resources for more severe delays. That is, a shared bus will not be dispatched to a route unless the expected delay is larger than a threshold. This approach has limitations as well. Similar to the FCFS strategy, dispatching decision is made regardless of the future state of the system.

To avoid these issues, the dispatching problem of the shared buses is formulated as an optimization problem. The schedule of a route is a sequence of departures associated with timestamps, referred to as "dispatching times". At each dispatching time, an available shared bus can potentially be dispatched in the absence of a dedicated bus. It is assumed that an AVL system is available providing real-time vehicle location information, which can help make more informed dispatching decisions. The decision to use a shared bus or save it for the future is formulated as an optimization problem that considers the locations of the nearest buses on all routes connected to the hub with the objective of minimizing delays. The formulation of the problem also considers the schedule of all the routes in the hub. It is assumed that the available fleet at the hub, as well as the $k$ nearest buses arriving at the hub on each route are known.

A mixed integer linear program is formulated to decide on whether to dispatch a bus from the shared fleet or not by minimizing some measures of cost (e.g., total passenger delay). The decision variable $x_{ij}$ assigns bus $i$ to trip $j$ and it is only defined when a bus to trip assignment is feasible. For example, a dedicated bus can only be assigned to trips of the route to which is dedicated, while shared buses can potentially be assigned to any route. If the optimization assigns a shared bus to the trip under consideration, the shared bus will be dispatched.

$$\text{Minimize } \sum_i \sum_j (C_{ij} x_{ij}) + \sum_j P_j (1 - \sum_i x_{ij})$$

s.t:

| | | |
|---|---|---|
| $\sum_j x_{ij} \leq 1$ | $j \in D$ and $x_{ij} \in \mathcal{M}$ | (6) |
| $\sum_i x_{ij} \leq 1$ | $i \in B$ and $x_{ij} \in \mathcal{M}$ | (7) |
| $x_{ij} = 0,1$ | | |

The objective function minimizes the total departure delay and includes a penalty term for missed trips (trips that are not assigned to any bus). $C_{ij}$ is the cost (delay) associated with assigning bus $i$ to trip $j$ (i.e. the difference between the scheduled time of trip $j$ and the time bus $i$ can be dispatched), and $P_j$ is the penalty term penalizing trips that have no bus assigned to them. Equation 6 guarantees that a bus is assigned to at most one trip. Equation 7 guarantees that a trip is assigned to at most one bus.

Figure 7 shows the setup of the problem. $B_s^h$ represents the set of shared buses that are either available at the hub or are arriving there. $B_r^h$ represents the set of buses that are dedicated to route $r$ that are available at the hub (these sets can be empty). $B_r$ is the set of dedicated buses on route $r$ on their way to the hub. For each route, the $k$ nearest buses arriving at the hub are considered, the shared buses (if any) belong to $B_s^h$ and the dedicated buses belong to set $B_r$. The $k$ immediate scheduled trips of each route are shown on the right side of Figure 7 and are represented by $D_r$. The lines connecting the circles show potential assignments of buses to trips. A blue line means that the buses are dedicated and can only serve trips of a specific route, while orange lines mean that buses are shared and can potentially serve trips on any route. As mentioned above, the dispatching problem is solved in real time, at every dispatching time.

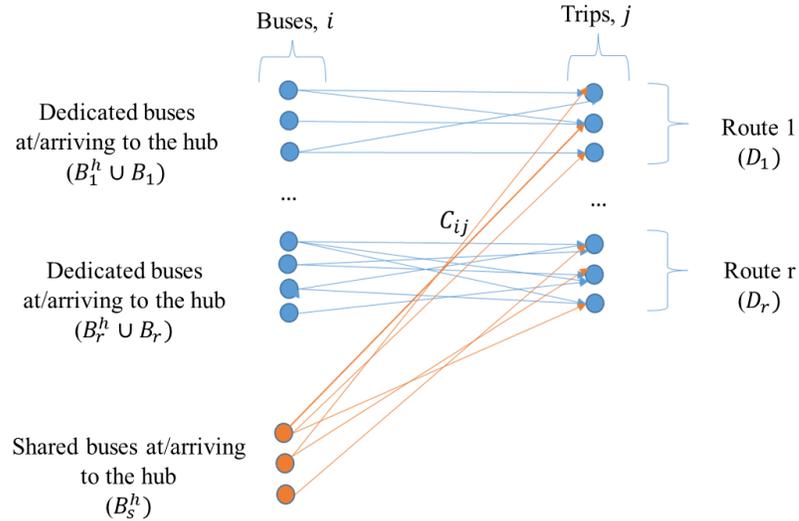

**Figure 7. Setup of the optimization problem. Dedicated buses can only be assigned to their corresponding routes (blue lines) while shared buses can be assigned to any route (orange lines).**

*Dispatching of dedicated buses*

The above section deals with the dispatching of shared buses. A heuristic rule-based strategy is used for the dispatching of dedicated buses under two cases: a) buses have schedules, and b) buses are schedule-free. In case (a), a dedicated bus performs a specific set of trips at specified times. If the bus arrives earlier than the departure time for the trip, it will wait until the scheduled departure time. On the other hand, if a bus is late, it will be immediately dispatched. In the second case, buses have no specific trips assigned to them and they can be dispatched anytime on their dedicated routes (this case is similar to the schedule-free operations, in (3). As a result, dispatching is based on the route timetable and the availability of dedicated buses. Once it is the time for a departure, all available dedicated buses for that route are considered. The earliest available dedicated buses is dispatched.

## 5. EVALUATION METHODOLOGY

A simulation model is developed to systematically examine the benefits of dynamic interlining. The model approximates the actual bus operations by assuming that buses operate as a shuttle traveling from/to a hub. Buses can have one or several checkpoints along the routes where they need to meet the schedule (other intermediate stops are not modeled). Each route has at least one of its terminal stations at a hub where it can use the shared fleet if needed. Terminal to terminal run times have two components: a) moving time and b) dwell time. Moving time pertains to the time that the bus is traveling between stops, and dwell time refers to the times the bus is spending picking up or dropping passengers at the stop. Dwell times are a function of actual headways and demand rates. The routes operate according to a pre-specified schedule. Buses are dispatched according to the procedure discussed in Section 4. The simulation considers various features such as the shared fleet size, run time variability, frequency of service, and etc., and outputs appropriate performance metrics to systematically evaluate dynamic interlining.

### 5.1 Simulation Model

Figure 8 illustrates the structure of the simulation model. The main inputs include the distribution of bus moving times, scheduled headways, ridership, fleet size, and network configuration. The network configuration determines how routes are aligned relative to each other, the number of check points on the routes, and the number of hubs in the network. The simulation consists of 4 modules: a) a dispatching module that governs the departure of buses from the check points, b) a fleet allocation module that assigns dedicated buses to their routes and shared buses to the shared fleet, c) a run times generator module that generates run times for the buses based on the state of the routes; and d) an output module that records bus movements in the network and calculates performance metrics.

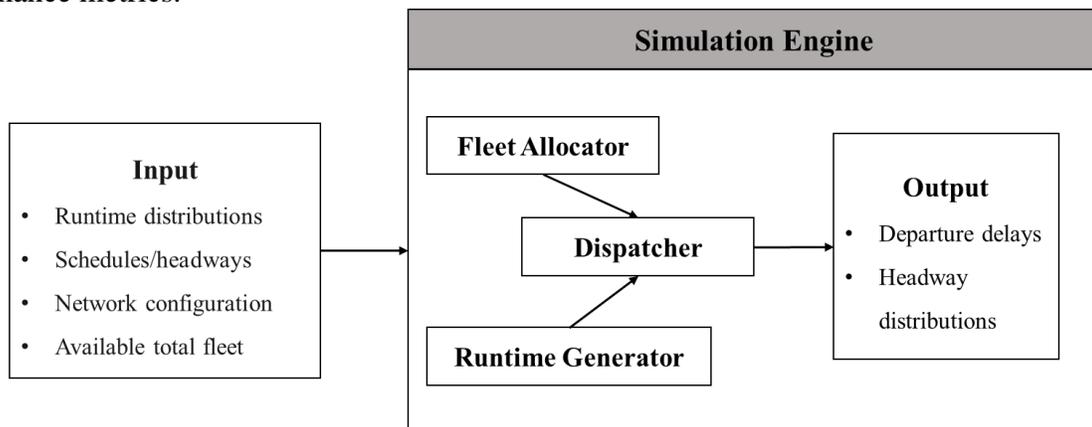

**Figure 8. Structure of the simulation model**

The simulation is time-based. At each simulation epoch, all routes are evaluated. If it is time to dispatch a bus on a route, the dispatcher selects an appropriate bus for the trip. Once a bus is chosen, the run time generator module generates the trip duration of that bus. The trip duration includes the vehicle moving time (randomly sampled from the moving time distribution) and the total dwell time (approximated using demand and departure headways). A dispatching error that captures delays due to human error and other random sources is also generated and impacts the actual departure time of a bus.

The moving time is the time that buses are actually moving between stops and can be impacted by weather, traffic conditions, driver behavior, and etc. The moving time is modeled as a random variable following a given distribution (theoretical or empirical) estimated from AVL.

The dwell time is route-level and calculated for each trip based on Sindzingre ( 2019) This model estimates the total dwell time of a bus trip as a function of the total number of passengers served by the bus and the number of stops made by that bus during the trip. The trip-level dwell time model in the simulation uses a similar idea and expresses the dwell time as a function of the number of passengers served. The number of passengers served by a trip is a function of the route's average ridership and the headway between two consecutive trips.

$$t_{r,i}^{dwell} = \alpha_r^{dwell} * h_i^r * Q_r + \beta \qquad (8)$$

where, $t_{r,i}^{dwell}$ is the trip-level dwell time expected for trip $i$ on route $r$, $\alpha_r^{dwell}$ is the average boarding time per passenger on route $r$, $h_i^r$ the departure headway between trips $i-1$ and $i$ on route $r$, $Q_r$ the average ridership per unit of time of route $r$, and $\beta$ the intercept.

In real operations, it may happen that a bus is available for an on-time departure, but the trip is delayed due to reasons such as driver breaks, dispatcher inattentiveness, unexpected dwell times at the dispatching station, and etc. This is referred to as "dispatching error" in the paper.

**5.2 Performance Metrics**

Three indicators are used to measure the effectiveness of dynamic interlining: the average departure delays from the terminal, the coefficient of variation of departure headways (CoV), and the ratio of the actual passenger wait times to the expected ones according to the schedules. The distribution of departure delays is a useful measure to compare the service that is delivered to what was planned. The coefficient of variation of headways is an important reliability measure as large variations in headways on high frequency routes result in increases in crowding and wait times (Pilachowski, 2009). Passenger wait times is an important metric of the level of service that a transit operation delivers. Since the passenger wait times are affected by service frequency and service reliability, the ratio of the experienced passenger wait times over the planned passenger wait times given the schedule is also used (scheduled wait time). Scheduled wait time is the expected wait time of passengers if all departures are on schedule. It can be estimated as ½ the scheduled headway.

**6. EMPIRICAL ANALYSIS**

An empirical analysis is conducted to evaluate and quantify the potential of dynamic interlining using realistic characteristics of bus operations

**6.1 Network and Data**
A terminal from the Massachusetts Bay Transit Authority (MBTA) systems, is used as the hub to evaluate the dynamic interlining strategy. The network consists of four relatively high frequency routes. The route operating characteristics were obtained from GTFS, AVL, and AFC data. The average route headways during the peak period (7:00 AM – 9:30 AM) were used. Two years of AVL data was used to estimate the moving time distributions. AFC data was used to estimate the ridership of the routes per direction. Table 2 summarizes the characteristics of the routes.

Table 2. Characteristics of the routes

| Route | Headway (seconds) | Average round trip run time (seconds) | Fleet size (buses) |
|---|---|---|---|
| A | 540 | 4761 | 10 |
| B | 510 | 6200 | 14 |
| C | 420 | 2972 | 9 |
| D | 360 | 2932 | 10 |

## 6.2 Experimental Design

The experiments are designed to explore the impact of the various factors on dynamic interlining performance, including fleet size, network, and service operating characteristics. Table 3 summarizes the factors investigated in the experiments.

Table 3. Factors investigated in the experiments.

| Category | Details | Description |
|---|---|---|
| Fleet Size | Shared fleet | Number of shared buses |
| | Dedicated fleet | Number of dedicated buses |
| Route Characteristics | Run time variability | CoV of run times of different routes |
| | Frequency | Frequency of service of different routes |
| Network | Number of routes | Number of routes that are dynamically interlined |
| | Within-hub travel times | Time it takes at the terminal for buses to switch to a different route |
| Operations | Bus passing | Whether or not buses can pass each other |
| | Bus schedules | Whether or not buses have schedules (or are schedule-free) |

A total of 59 scenarios were tested. Each scenario aims at evaluating combinations of the factors listed in Table 3. The base case scenario represents, to the extent possible, the operating characteristics of the routes without dynamic interlining. In the base case scenario, all buses are dedicated and have schedules, there is no dynamic interlining, buses can pass each other, and the characteristics of the routes are the same as in Table 2.

## 6.4 Results

*Overall potential*

Figure 9 compares system performance for the two extreme operating scenarios under different fleet sizes: a scenario with scheduled buses that are assigned to specific trips; and a scenario where all buses are shared and dynamically interlined. The results show that dynamic interlining not only provides better reliability for a given fleet size but also can reduce the required fleet size for a target reliability. In other words, by switching from operations with buses having schedules to dynamic interlining, similar or better reliability can be achieved while reducing the required fleet size. For example, current operations (base case) utilize a total of 43 buses and users experience an average wait time of 252 seconds. When the system operates with all buses shared, they experience similar wait times with only 39 buses. The corresponding departure delays are even less.

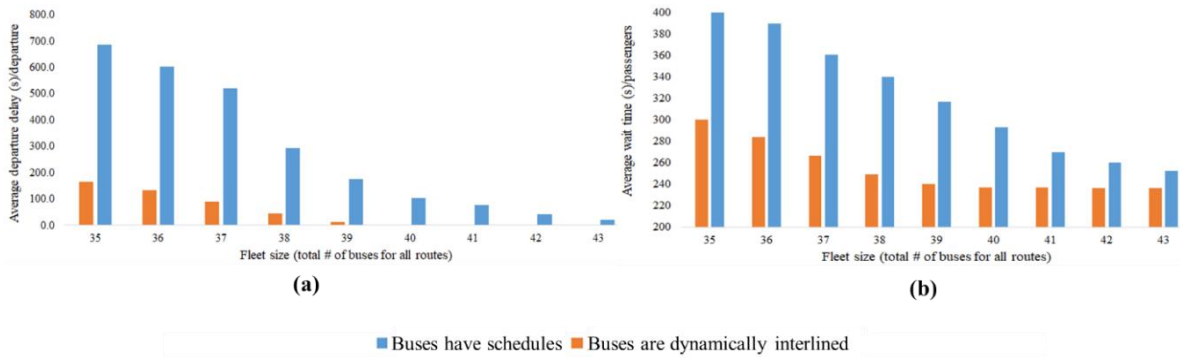

**Figure 9. Comparison of departure delays (a) and wait times (b) for dynamic interlining and scheduled buses for different fleet sizes**

*Impact of shared fleet size*

Figure 10 shows the coefficient of variation (CoV) of for different shared fleet sizes. Each bar shows the CoV of headways of the corresponding routes for a scenario. The darker the color of the bars the larger the size of the shared fleet. Zero means that although buses are not shared among routes, they are not scheduled for specific trips either, representing the case of within route interlining, similar to schedule-free operations (Section 4.2). The red arrows on top of the bars indicate the route that contributes a bus so that to the shared fleet increases by 1. This coincides with an increase in the headway CoV for that route. However, it is worth noting that this increase becomes smaller as the number of shared buses subsequently increases as other routes contribute shared buses. This is the price the route pays for the system improvement. The overall trend indicates that the reliability increases as more buses are shared. Except for the slight increase in the CoV of the contributing route, the CoVs decrease when more buses are dynamically interlined. In the case with all buses dynamically interlined, the performance of all routes significantly improves.

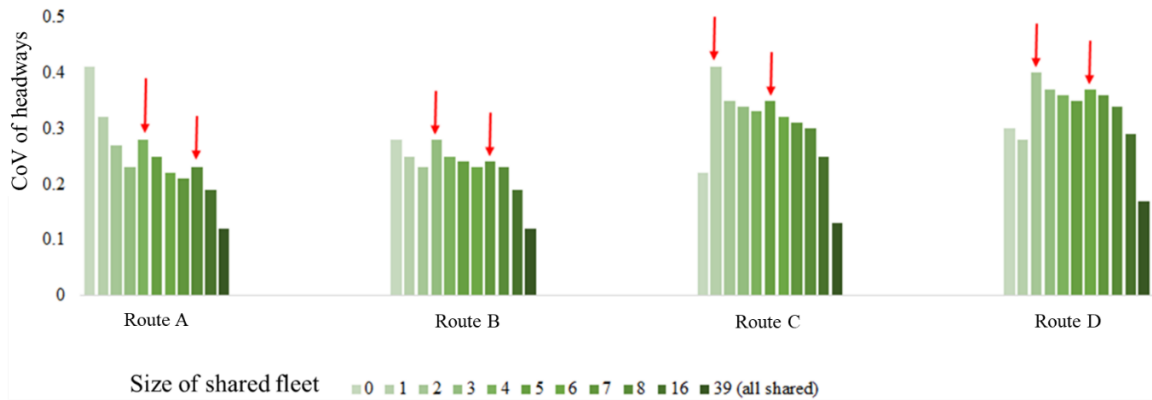

**Figure 10 Impact of shared fleet on route performance for different shared fleet sizes**

Figure 11 illustrates the average departure delay per departure as a function of different shared fleet sizes. The performance slightly deteriorates when only very few buses are shared, however, the delays keep decreasing as more buses are added to the shared fleet since the opportunities to dynamically interline the buses increase. Moreover, bus idle times also decrease from 334 seconds per departure when fleet is dedicated to 258 when all buses are shared.

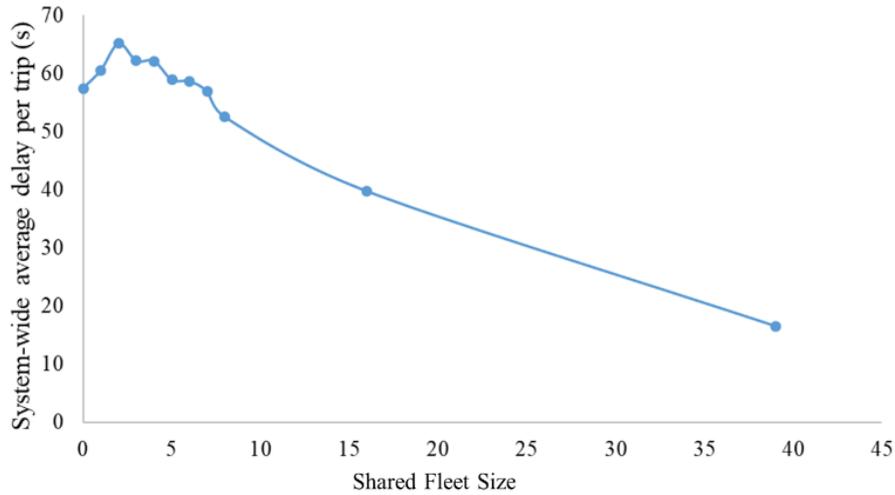

**Figure 11. Average system-wide departure delays as a function of shared fleet size**

*Impact of travel time within the hub*

In practice, the terminals of the routes may be located at some distance from each other within the hub. Table 4 shows the impact of within-hub travel times on departure delays. A small within-hub travel time (<5 minutes) can be absorbed by dynamic interlining and it only slightly affects performance. However, as intra-terminal distances increase, the benefit of dynamic interlining disappears, and the optimal dispatching strategy does not favor the use of shared buses on alternative routes. When the inter-terminal travel times within the hub are 10 minutes, the dynamic interlining solution is the same as the scheduled free case.

**Table 4. Impacts of travel times within a hub on dynamic interlining**

| Type of operation | Travel time within hub (min) | Average delay (s) | Average wait time (s) |
|---|---|---|---|
| Buses have schedules (base case) | - | 175 | 317 |
| Buses are not shared and are schedule-free | - | 57 | 254 |
| All dynamically interlined | 0 | 16 | 240 |
| | 2 | 19 | 243 |
| | 5 | 32 | 248 |
| | 10 | 57 | 254 |

*Impact of other factors*

Table 5 summarizes the results for the remaining scenarios. Each row shows the impact of dynamic interlining on the performance of the system under various operating scenarios. Performance is measured as the percent reduction in average departure delay per departure and wait time ratios (actual wait time to expected wait time based on the scheduled headways) in the system because of interlining compared to the case of no interlining. The scenarios assume that the base case fleet is 39 and in the case of dynamic interlining all buses are shared.

**Table 5. Impact of route/operating characteristics on dynamic interlining performance (39 buses)**

| Scenario | Percent reduction when dynamic interlining is used relative to each scenario | |
|---|---|---|
| | Delay | Wait time ratio |
| Base case (39 buses) | 91% | 23% |
| Buses not allowed to pass each other | 69% | 24% |
| Frequency doubled | 98% | 39% |
| # of routes doubled | 97% | 25% |
| CoV of run times doubled | 92% | 29% |

- *Bus passing constraints*. Some operators do not allow buses to pass each other and would require the trailing buses to slow down. The results indicate that dynamic interlining is more effective when buses are allowed to pass each other, since it increases the opportunities for interlining of the buses.
- *Frequency of service*. To investigate the effects of frequency on dynamic interlining, frequencies of all routes are increased by 100% (fleet sizes are adjusted accordingly). The results show that dynamic interlining is more effective when routes have higher frequencies. For example, the reduction in delays for the cases with the original frequencies, and doubled frequencies is 91% and 98% respectively relative to the operation without dynamic interlining.
- *Number of routes*. To examine the impact of the number of routes that participate in dynamic interlining, routes identical to the routes in the base case are added to the hub doubling the number of the routes. The results show that the delays per departure decrease as more routes are involved in dynamic interlining. As with increased frequency of service, increasing the number of routes improves the dynamic interlining performance. In both cases, the fleet size is increased which means more buses are shared among the routes.
- *Uncertainty in run times*. Dynamic interlining performance under larger coefficients of variation of run times is evaluated (CoV of run times were increased by 100%). The fleet sizes are kept the same as the base case for these scenarios. The results show that dynamic interlining improves reliability when run time variability is higher. Conceptually, dynamic interlining should be more effective when there is more uncertainty in bus run times. In the extreme case when there is no variation in run times, the dynamic interlining and scheduled buses strategies are equivalent.

## CONCLUSION

The paper proposes a dynamic interlining strategy in bus operations and explores its performance. Dynamic interlining reserves a portion of (or all) the fleet as a shared fleet and allows the routes to use them on an on-demand basis aiming at reducing delays and regulating service. A simulation model is developed to systematically examine the impact of dynamic interlining on service reliability and explore the key factors that affect its performance. The simulation model considers the stochasticity of bus operations by incorporating run time distributions, dispatching errors, and additional dwell times caused by uneven headways. The bus dispatching problem in the case of dedicated and shared buses is formatted as an optimization problem and used by the simulation to optimally allocate shared buses to routes.

The analysis shows that dynamic interlining can improve service reliability compared to scheduled operations. The benefits increase as more buses are shared and as variability in run times increases. Allowing buses to pass each other, high-frequency routes, and having more routes participate in the strategy, increase the effectiveness of dynamic interlining. However, inter-terminal travel times within the hub reduce the effectiveness of dynamic interlining.